%Paper: hep-th/9404009
%From: jspark@phya.yonsei.ac.kr (Jae-Suk Park)
%Date: Sun, 3 Apr 1994 22:56:20 +0900 (KST)
%Date (revised): Mon, 4 Apr 1994 01:27:55 +0900 (KST)
%Date (revised): Fri, 9 Sep 1994 18:28:44 +0900 (KDT)

\input harvmac

%----------------------------------------------------------------------------
%You need AMSfonts 2.1. If you do not have
%AMSfonts, please search for the word "Euler" and follow the instructions.
%----------------------------------------------------------------------------
%----------------------------------------------------------------------------
% personal macros
%----------------------------------------------------------------------------
\def\title#1#2#3#4{\nopagenumbers\abstractfont\hsize=\hstitle
\rightline{#1}%
\rightline{#2}
\bigskip
\vskip 0.7in\centerline{\titlefont #3}\abstractfont\vskip .2in
\centerline{{\titlefont#4}}
\vskip 0.3in\pageno=0}
\def\ack{\bigbreak\bigskip\bigskip\centerline{{\bf Acknowledgements}}\nobreak}
\baselineskip=14pt plus 2pt minus 1pt
\newskip\normalparskip
\normalparskip = 6pt plus 2pt minus 1pt
\parskip = \normalparskip
\parindent=16pt
\def\submit{\baselineskip=20pt plus 2pt minus 2pt}
%----------------------------------------------------------------------------
%symbols
%----------------------------------------------------------------------------
\def\a{\alpha}    \def\b{\beta}       \def\c{\chi}       \def\d{\delta}
\def\D{\Delta}    \def\e{\varepsilon}        \def\F{\Phi}
\def\g{\gamma}    \def\G{\Gamma}           
\def\L{\Lambda}                    \def\r{\rho}
  \def\o{\omega}      \def\O{\Omega}     \def\p{\psi}
\def\P{\Psi}      \def\s{\sigma}      \def\S{\Sigma}     
        \def\w{\varphi}    

\def\CA{{\cal A}}
\def\CB{{\cal B}}
\def\CG{{\cal G}}
\def\CM{{\cal M}}
\def\CE{{\cal E}}
\def\CL{{\cal L}}
\def\CD{{\cal D}}

\def\CQ{{\cal Q}}
%
%------------------------------------------------
% Euler Fonts
\font\teneufm=eufm10
\font\seveneufm=eufm7
\font\fiveeufm=eufm5
\newfam\eufmfam
\textfont\eufmfam=\teneufm
\scriptfont\eufmfam=\seveneufm
\scriptscriptfont\eufmfam=\fiveeufm
\def\eufm#1{{\fam\eufmfam\relax#1}}

\font\teneusm=eusm10
\font\seveneusm=eusm7
\font\fiveeusm=eusm5
\newfam\eusmfam
\textfont\eusmfam=\teneusm
\scriptfont\eusmfam=\seveneusm
\scriptscriptfont\eusmfam=\fiveeusm

\font\tenmsx=msam10
\font\sevenmsx=msam7
\font\fivemsx=msam5
\font\tenmsy=msbm10
\font\sevenmsy=msbm7
\font\fivemsy=msbm5
\newfam\msafam
\newfam\msbfam
\textfont\msafam=\tenmsx  \scriptfont\msafam=\sevenmsx
  \scriptscriptfont\msafam=\fivemsx
\textfont\msbfam=\tenmsy  \scriptfont\msbfam=\sevenmsy
  \scriptscriptfont\msbfam=\fivemsy

\def\msbm#1{{\fam\msbfam\relax#1}}

%\def\eufm#1{{\bf #1}}  %If you do not have AMSfonts 2.1,
%\def\eusm#1{{\cal #1}} %uncomment these three lines
%\def\msbm#1{{\bf #1}}  %and comment the lines below %Euler Fonts

%--------------------------------------------------------------------------
%Math
\def\rd{\partial}

\def\darr#1{\raise1.5ex\hbox{$\leftrightarrow$}\mkern-16.5mu #1}
\def\Ha{{1\over2}}
\def\ha{{\textstyle{1\over2}}}
\def\fr#1#2{{\textstyle{#1\over#2}}}
\def\Fr#1#2{{#1\over#2}}
\def\tr{\hbox{Tr}\,}

   %Feynman dagger
 %Feynman dagger
\def\roughly#1{\raise.3ex\hbox{$#1$\kern-.75em\lower1ex\hbox{$\sim$}}}

%
%Journals
\def\cmp#1#2#3{Comm.\ Math.\ Phys.\ {{\bf #1}} {(#2)} {#3}}

\def\np#1#2#3{Nucl.\ Phys.\ {{\bf #1}} {(#2)} {#3}}

\def\jdg#1#2#3{J.\ Differ.\ Geom.\ {{\bf #1}} {(#2)} {#3}}

\def\top#1#2#3{Topology {{\bf #1}} {(#2)} {#3}}
\def\zp#1#2#3{Z.\ Phys.\ {{\bf #1}} {(#2)} {#3}}

\def\im#1#2#3{Invent.\ Math.\ {{\bf #1}} {(#2)} {#3}}

\def\bams#1#2#3{Bull.\ Am.\ Math.\ Soc.\ {{\bf #1}} {(#2)} {#3}}
\def\jgp#1#2#3{J.\ Geom.\ Phys.\ {{\bf #1}} {(#2)} {#3}}
%---------------------------------------------------------------------
%Macros specific to this file

\def\pr{\prime}
\def\ppr{{\prime\prime}}

\def\bs{{\bf s}}
\def\bbs{{\bf\bar s}}
\def\Da{d_{\!A}}
\def\Dp{\rd_{\!A}}

\def\Dpp{\bar\rd_{\!A}}

\def\gE{\eufm{g}_{\raise-.1ex\hbox{${}_E$}}}
\def\gBE{\eufm{g}_{\raise-.1ex\hbox{${}_\BE$}}}
\def\gEc{\eufm{g}_{\raise-.1ex\hbox{${}_E$}}^C}
\def\BC{\msbm{C}}
\def\BE{\msbm{E}}

\def\BZ{\msbm{Z}}

\def\dw{\d_{\!{}_{W}}}

\def\rem#1{\medskip\leftline{${\hbox{{\it #1}}}$}}

%-------------------------------------------------------------------
\lref\WittenA{
E.~Witten,
Topological quantum field theory,
\cmp{117}{1988}{353}.
}
\lref\WittenB{
E.~Witten,
Supersymmetric Yang-Mills theory on a four manifolds,
Preprint IASSNS-HEP-94/5 February, 1994.
%hep-th/9403195.
}
\lref\WittenC{
E.~Witten,
Two dimensional gauge theories revisited.
\jgp{9}{1992}{303}.
}
\lref\WittenD{
E.~Witten,
Mirror manifolds and topological field theory,
in: Essays on Mirror Manifolds, ed.~S.-T.~Yau,
(International Press, Hongkong, 1992)
%IASSNS-HEP-91/83 \& hep-th/9112056.
}
\lref\WittenE{
E.~Witten,
The $N$ matrix model and gauged WZW models,
\np {B 371}{1992}{191}.
}
\lref\WittenP{
E.~Witten, private communication.
}
\lref\DonaldsonA{
S.K.\ Donaldson, Polynomial invariants for smooth $4$-manifolds,
\top{29}{1990}{257}.
}
\lref\ParkA{
J.-S.\ Park,
$N=2$ topological Yang-Mills theory on compact
K\"{a}hler surfaces, \cmp{163}{1994}{113}
}
\lref\ParkB{
J.-S.\ Park, Holomorphic Yang-Mills theory on compact
K\"{a}hler Manifolds, \np {B423}{1994}{559}
}
\lref\HPb{
S.J.~Hyun and J.-S.~Park,
A Dolbeault model for equivariant cohomology and equivariant
differential forms,
in preparation.
}
\lref\HPa{
S.J.~Hyun and J.-S.~Park,
$N=2$ Topological Yang-Mills Theory and
Donaldson's polynomials II,
in progress.
}
\lref\KM{
P.~Kronheimer and T.~Mrowka, Recurrence relations and asymptotics
for four-manifold invariants, \bams{30}{1994}{215}.
}
\lref\BrA{
R.~Brussee, On the $(-1)$-curve conjecture of Friedman and Morgan,
\im{114}{1993}{219}.
}
\lref\BrB{
R.~Brussee, Some remarks on the Kronheimer-Mrowka classes of
algebraic surfaces, preprint alg-geom/9308003 .
}
\lref\JL{
J.~Li, Algebraic geometric interpretation of Donaldson's polynomial
invariants, \jdg{37}{1993}{417}.
}
\lref\BS{
L.\ Baulieu and I.M.\ Singer,
Topological Yang-Mills symmetry,
\np{(Proc.\ Suppl.) 5B}{1988}{12}.
}
\lref\Kanno{
H.\ Kanno,
Weil algebraic structure and geometrical meaning of the BRST
transformation in topological quantum field theory,
\zp{C 43}{1989}{477}.
}
\lref\AJ{
M.F.~Atiyah and L.~Jeffrey,
Topological Lagrangians and cohomology,
\jgp{7}{1990}{1}.
}

\lref\AB{
M.F.~Atiyah and R.~Bott,
The moment map and equivariant cohomology,
\top{23}{1984}{1}
}
\lref\Kalk{
J.~Kalkman,
BRST model for equivariant cohomology and representatives for the
equivariant Thom class,
\cmp{153}{1993}{447}
}
\lref\KalkC{
J.~Kalkman,
Residues in nonabelian localization,
hep-th/9406
}
\lref\Wu{
S.~Wu, An integration formula for the square of moment maps of circle
actions, to appear in Lett.~Math.~Phys.
}
\lref\JK{
L.C.~Jeffrey and F.C.~Kirwan,
Localization for non-abelian group actions, alg-geom/9307001
}
\lref\PW{
E.~Prato and S.~Wu, Duistermaat-Heckman measures in a non-compact
setting, alg-geom/9307005
}
\lref\Verg{
M.~Vergne,
A note on Jeffrey-Kirwan-Witten's localization formula, preprint LMENS -94-12
}
\lref\MQ{
V.~Mathai and D.~Quillen,
Thom classes, superconnections and equivariant diffrential forms,
\top{25}{1986}{85}
}
\lref\Grady{
K.G.~O'Grady,
Donaldson's polynomials for $K3$ surfaces,
\jdg{35}{1992}{415}
}
\lref\DK{
S.K.\ Donaldson and P.B.\ Kronheimer,
The geometry of four-manifolds (Oxford University Press, New York, 1990)
}
\lref\Kirwan{
F.C~Kirwan,
Cohomology of quotients in algebraic and symplectic geometry,
(Princeton Univ.~Press, Princeton, NJ, 1985)
}
\lref\Tuyrin{
A.N.~Tyurin,
Algebraic geometric aspects of smooth structure 1. The Donaldson
Polynomials, Uspekhi Mat.~Nauk {\bf 44}:3 (1989) 93 (Russ.~
Math.~Surv.~{\bf 44}:3 (1989) 113)
}
\lref\Kobayashi{
S.\ Kobayashi,
Differential geometry of complex vector bundle
(Princeton University Press, Princeton, 1987)
}
%---------------------------------------------------------------------
\title{YUMS-94-08[revised]}{hep-th/9404009[revised]}
{$N=2$ Topological Yang-Mills Theories}
{and Donaldson's Polynomials}

\bigskip
\bigskip
\centerline{
Seungjoon Hyun\footnote{$^{\dagger}$}{e-mail: hyun@phya.yonsei.ac.kr}
and
Jae-Suk Park\footnote{$^{\dagger\dagger}$}{address after Oct.~1994:
Theoretical Physics Group, The University College of Swansea, UK.}
}
\bigskip
\bigskip
\centerline{{\it Institute for Mathematical Science}}
\centerline{{\it Yonsei University}}
\centerline{{\it Seoul 120-749}}
\centerline{{\it Korea}}
\bigskip\bigskip \bigskip
\bigskip
%abstract
\centerline{
{\bf Abstract}
}
\vskip 0.2in
The $N=2$ topological Yang-Mills and holomorphic
Yang-Mills theories on simply connected compact K\"{a}hler
surfaces with $p_g\geq 1$ are reexamined.
The $N=2$ symmetry is clarified in terms of
a Dolbeault model of the equivariant cohomology.
We realize the non-algebraic part of Donaldson's polynomial
invariants as well as the algebraic part.
We calculate Donaldson's polynomials on
$H^{2,0}(S,\BZ)\oplus H^{0,2}(S,\BZ)$.
\bigskip
\bigskip

\centerline{1991 MSC: 81 T 13, 81 T 60, 55 N 91, 53 C 07}

\Date{March, 1994, Revised by August, 1994}
%\draftmode
\submit
\newsec{Introduction}

The $N=2$ super Yang-Mills theory on arbitrary four-manifolds
can be twisted to define $N=1$ topological Yang-Mills (TYM) theory
which realize Donaldson's polynomial invariants of smooth four-manifolds
\DonaldsonA\DK\ as correlation functions \WittenA.
Recently, Witten has determined the Donaldson invariants of compact
K\"{a}hler surfaces with $p_g\geq 1$ by exploiting some standard
properties of $N=2$ and $N=1$ super Yang-Mills theories \WittenB.

Some time ago,  the second author proposed $N=2$ TYM theory on
compact K\"{a}hler surfaces \ParkA. His construction is based directly
on the $N=1$ TYM theory utilizing the complex and
K\"{a}hler structures of the moduli space of anti-self-dual (ASD)
connections. He has also proposed $N=2$ holomorphic Yang-Mills (HYM) theory
whose partition function is a generating functional of certain Donaldson
invariants \ParkB, adapting the two-dimensional construction of Witten's
to K\"{a}hler surfaces \WittenC. However, both
theories describe the algebraic part of the Donaldson invariants,
analogous to the invariants defined by J.~Li \JL, rather than all the
invariants and the non-algebraic part was simply
ignored. Futhermore, we will see that it is impossible to realize the
non-algebraic part in those constructions. The purpose of this paper is
to fill those gaps.
%We were able to realize this after reading \WittenB.

In this paper, we reexamine $N=2$ TYM and HYM theories
on simply connected compact K\"{a}hler surfaces with $p_g\geq 1$,
which lead to the different $N=2$ (global) supersymmetry
transformation laws for some auxiliary fields.
This allows us to realize the
non-algebraic part of Donaldson's polynomials as well as the
algebraic part. We calculate Donaldson's polynomial invariants on
$H^{2,0}(X,\BZ)\oplus H^{0,2}(X,\BZ)$.

This paper is organized as follows; in sect.~2, we
give backgrounds and motivations of this paper.
We compare the basic supersymmetry transformation
laws of the $N=1$ and the $N=2$ TYM theories in terms of
the de Rham and a Dolbeault models of the equivariant cohomology.
We show that the Dolbeault equivariant cohomology is not
isomorphic to the de Rham equivariant cohomology.
In the field theoretical context, this amounts to
introducing on-shell observables in the $N=2$ TYM and HYM theories.
In sect.~3, we construct new $N=2$ TYM theory. We briefly discuss
the geometrical and the physical meanings of fermionic zero-modes.
We resolve the problem of the on-shell invariants adapting Witten's
method of introducing the mass gap \WittenB.
In sect.~4, we study deformations to HYM theories and calculate
Donaldson's invariants on $H^{2,0}(X,\BZ)\oplus H^{0,2}(X,\BZ)$.
We also show that the broken part of the $N=2$ supersymmetry
due to the mass gap is restored in the prcess of the  deformation.
We compare our results with others and give some general remarks on
the algebraic parts of the invariants. Our method will lead us to
determine the full invariants for simply connected $K3$ surfaces.

The algebraic part of Donaldson's polynomials will be studied in our
future publication \HPa.

\newsec{Backgrounds and Motivations}

We consider a simply connected compact K\"{a}hler surface
$X$ with K\"{a}hler form $\o$ and $b_2^+ = 1 + 2p_g \geq 3$
where $b_2^+$ and $p_g$ denote the number of the self-dual harmornic two
forms and the geometric genus, respectively.
Let $E$ be a complex vector bundle over $X$ with
the restriction of structure group to $SU(2)$.
We write $\gE$ for the Lie algebra bundle associated with $E$ by
adjoint representation. We introduce a positive definite quadratic
form $(a,b) = -\tr ab$ on $\eufm{su(2)}$, where $\tr$ denotes the
trace in the $2$-dimensional representation. Then, the bundle $E$
is classified by the instanton number;
$$
k =<c_2(E),X>=\Fr{1}{8\pi^2}\int_X \tr F\wedge F \in \BZ.
$$

Let $\CA$ denote the space of
all connections, which is an affine space whose tangent
vectors are represented by $\gE$-valued one form $\d A \in \O^1(\gE)$.
Let $\CG$ be the group of gauge transformations.
%For our purpose, there is no loss of generality to consider the space
%of all irreducible connections instead of the space $\CA$ and we will
%abuse notations

\subsec{The $N=1$ supersymmetry}

The global supersymmetry operator $\dw$ of the $N=1$ topological Yang-Mills
theory can be interpreted as the exterior (covariant) derivative on
$\CA/\CG$ \WittenA\Kanno. The $N=1$ supersymmetry transformation
laws for the basic multiplet $(A,\P,\F)$ are
\eqn\waa{
\dw A =-\P,\qquad
\dw \P = -i\Da\F,\qquad
\dw \F = 0,
}
where $\P \in \O^{1}(\gE)$ and $\F \in
\O^{0}(\gE)$. One introduce a global quantum number (or the ghost number)
$U$ which assign the value $1$ to $\dw$. The $U$ numbers of the basic
fields $(A,\P,\F)$ are $(0,1,2)$. Note that $\dw^2 = -i\d_\F$,
where $\d_\F$ is the generator of a gauge transformation with
infinitesimal parameter $\F$. Thus, $\dw^2 = 0$ if it acts on a
$\CG$-invariant functional of the basic fields. The supersymmetry
operator $\dw$ can be viewed as the de Rham cohomology operator
on $\CA/\CG$ if $\CG$ acts freely on $\CA$.

More precisely, $\dw$ is the operator of the de Rham model for
the $\CG$-equivariant cohomology of \foot{
We refer the reader to \AB\WittenC\ for details. We generally
follows \WittenC.} $\CA$.
Let $Lie(\CG)$ be the Lie algebra of $\CG$ which is the space
$\O^0(\gE)$ of $\gE$-valued zero-forms.  The $\CG$ action
on $\CA$ is generated by vector fields $V_a$, where
we pick an orthonomal basis $T_a$ of $Lie(\CG)$.
Let $\hbox{Fun($Lie(\CG)$)}$ be the algebra  of polynomial functions,
generated by $\F^a$ with degree $2$, on $Lie(\CG)$.
The $\CG$-equivariant de Rham complex is
$\O^*_\CG(\CA) = (\O^*(\CA)\otimes \hbox{Fun}(Lie(\CG)))^\CG$.
The associated differential operator is $\dw$ which can be  represented
as
\eqn\wab{
\dw = -\sum_I \P^I\Fr{\rd}{\rd A^{I}} +i\sum_{I,a}\w^a
V^{I}_a \Fr{\rd}{\rd \P^{I}},
}
where $A^I$ are the local coordinates on $\CA$.
We have
\eqn\wac{
\dw^2 = -i\F^a\CL_a,
}
where  $\CL_a$ is the Lie derivative
with respect to $V_a$. Thus, $\dw^2 = 0$ on the $\CG$-invariant subspace
$\O^*_\CG(\CA)$ of $\O^*(\CA)\otimes \hbox{Fun}(Lie(\CG))$.
The $\CG$-equivariant de Rham cohomology $H^*_\CG(\CA)$ is defined as
the pairs $(\O^*_\CG(\CA),\;\dw)$.

In Donaldson-Witten theory, we are interested in the
$\CG$-equivariant cohomology of the space of anti-self-dual (ASD)
connections. Since there are no reducible ASD connections, for generic
metrics on $X$, $\CG$ acts freely on the space of ASD connections,
the $\CG$-equivariant cohomology reduces to the de Rham cohomology
of the moduli space $\CM$ of ASD connections. The de Rham cohomology
on $\CM$ can be obtained from $H^*_\CG(\CA)$ by restriction and
reduction. For example, an element of $H^2_\CG(\CA)$ is given by
\eqn\waf{
\tilde\o^{(2)}=\Fr{1}{8\pi^2}\int_X\tr(\P\wedge\P + i\F F_A)\wedge
\o^{(2)},
}
where $\o^{(2)} \in H^2(X,\BZ)$. The cohmology class of
$H^2_\CG(\CA)$ depends only on the cohomology class of
$H^2(X,\BZ)$. In Witten's approach, an element of $H^2(\CM)$
can be obtained from $H^2_\CG(\CA)$ by the field theoretical
methods, in which $\P$ is eventually replaced by its zero-modes,
$A$ by the ASD connection
and $\F$ by its vacumm expectation value.
One can also view $\tilde\o^{(2)}$ as the equivariantly closed
extension \AB\ of a closed form
$\Fr{1}{8\pi^2}\int_X\tr(\P\wedge\P)\wedge\o^{(2)}$ on $\CA$.
It is also known that any element of $H^*(\CM)$ is induced
from an element of $H^*_\CG(\CA)$ \DK\Kirwan.

\subsec{The $N=2$ supersymmetry}

Picking a complex structure $J$ on $X$, one
can introduce a complex structure $J_{\!\CA}$ on $\CA$ as well
as on $\CA/\CG$,
\eqn\cmme{
J_{\!\CA} \d A = J \d A\;,\qquad \d A \in T\CA\;,
}
by identifying $T^{1,0}\CA$ and $T^{0,1}\CA$ in $T\CA =
T^{1,0}\CA\oplus T^{0,1}\CA$ with the $\gE$ valued $(1,0)$-forms
and $(0,1)$-forms on $X$, respectively.
We can also introduce natural K\"{a}hler structure on $\CA$ with
K\"{a}hler form
\eqn\wad{
\tilde\o = \Fr{1}{8\pi^2}\int_X \tr (\d A\wedge \d A)\wedge \o.
}

Using the complex structures $J$ and
$J_{\!\CA}$, we can decompose $\dw = \bs +\bbs$ and
find the $N=2$ transformation laws
for the basic multiplet $(A^\pr,A^\ppr,\p,\bar\p,\w)$ \ParkA;
\eqn\symphony{\eqalign{
&\bs A^\pr =-\p,\cr
&\bbs A^\pr =0,\cr
&\bs A^\ppr =0,\cr
&\bbs A^\ppr =-\bar\p,\cr
}\qquad\eqalign{
&\bs \p = 0,\cr
&\bbs\p = -i\Dp\w,\cr
&\bs\bar\p = -i\Dpp\w,\cr
&\bbs \bar\p = 0,\cr
}\qquad\eqalign{
\bbs\w=0,\cr
\bs\w=0,\cr
}}
where $\p \in \O^{1,0}(\gE)$, $\bar\p \in \O^{0,1}(\gE)$ and $\w \in
\O^{0,0}(\gE)$. Note that $\p$ can be identified with holomorphic
(co)tangent vectors on $\CA$. It is important to note that $\w$ is
of degree $(1,1)$.
We introduce two global quantum numbers (or ghost
numbers) $(U,R)$, which assign $(1,1)$ to $\bs$ and $(1,-1)$ to $\bbs$.
A quantity of degree $(p,q)$ has $U = p+q$ and $R = p-q$.
The above transformation laws play a central role in constructing
$N=2$ TYM theories.

The commutation relations of the fermionic symmetry generators
$\bs,\bbs$ are
\eqn\kkg{
\bs^2=0,\qquad (\bs\bbs +\bbs\bs) = i\Da \w = -i\d_\w,\qquad \bbs^2=0,
}
where $\d_\w$ is the generator of a gauge transformation with
infinitesimal parameter $\w$.
Thus $\{\bs,\bbs\} = 0$ precisely
on the $\CG$ invariant space or if it acts on $\CG$-invariant
functionals of $A^\pr, A^\ppr,\p,\bar\p,\w$. Thus, $\bbs$
can be roughly viewed as the operator of Dolbeault cohomology group
on $\CA/\CG$.

In fact, $\bbs$ is the operator of
a Dolbeault cohomological analogue of $\CG$-equivariant cohomology\foot{
Note that the transformations \symphony\ are slightly different,
{\it in conventions}, from those in \ParkA. Here we follow the usual
conventions in physics literatures.
}
of $\CA$.
This can be formally described as follows;
we let $\O^{*,*}(\CA)$ be the Dolbeault complex on $\CA$.
Now we interpret $\hbox{Fun($Lie(\CG)$)}$ to the algebra of
polynomials functions generated by $\w^a$.
Then the desired Dolbeault model of the $\CG$-equivariant
complex is $\O^{*,*}_\CG = (\O^{*,*}(\CA)\otimes \hbox{Fun}(\CG))^\CG$.
The associated differential operators with the degrees $(1,0)$ and
$(0,1)$ are $\bs$ and $\bbs$ represented by
\eqn\equi{\eqalign{
&\bs = -\sum_i \p^i\Fr{\rd}{\rd A^{\pr i}} +i\sum_{\bar i,a}\w^a
V^{\bar i}_a \Fr{\rd}{\rd \bar\p^{\bar i}},\cr
&\bbs = -\sum_{\bar i} \bar\p^{\bar i}\Fr{\rd}{\rd A^{\ppr\bar i}}
+i\sum_{i,a}\w^a
V^{i}_a \Fr{\rd}{\rd \p^{i}},\cr
}}
where $i,\bar i$ are the local holomorphic and anti-holomorphic
indices tangent to $\CA$. We have
\eqn\Lie{
\bs^2=0,\qquad \bs\bbs + \bbs\bs = -i\w^a\CL_a,\qquad \bbs^2=0,
}
Thus, $\{\bs,\bbs\} = 0$ on the $\CG$-invariant subspace
$\O^{*,*}_\CG$ of $\O^{*,*}(\CA)\otimes \hbox{Fun}(Lie(\CG))$.
{\it We define the $\CG$-equivariant Dolbeault cohomology $H^{*,*}_\CG(\CA)$
by
the pairs $(\O^{*,*}_\CG(\CA),\;\bbs)$.}

An immidiate observation is that the analogue of the
Hodge decomposition theorem  will not be applicable in the equivariant
sense. Since $\CA$ has the K\"{a}hler structure, the de Rham and the
Dolbeault cohomologies on $\CA$ are related by the Hodge decompositions.
If we assume $\CG$ acts freely on $\CA$, we can expect our equivariant
Dolbealut cohomology $\O^{*,*}_\CG(\CA)$ is isomorphic to the usual
Dolbealut cohomology on $\CA/\CG$.
And, the equivariant de Rham cohomology $\O^*_\CG(\CA)$
is isomorphic to the de Rham cohomology of $\CA/\CG$.
Since the K\"{a}hler structure on $\CA$ does not descend
to $\CA/\CG$ in general, the Hodge decomposition theorem
is not applicable in general. That is,
a $\CG$-invariant  and $\bbs$-closed quantity is
not automatically $\bs$-closed one.

\subsec{The old construction}

In the old construction \ParkA, we
introduced an anti-ghost $B$, a self-dual two form
$B = B^{2,0} + B^{0,2} +B^0\o\in \O^2_+(\gE)$
in the adjoint representation, with
$(U,R) = (-2,0)$.
Then \kkg\ naturally leads us to
the multiplet $(B,i\c, -i\bar\c, H)$ with
transformation laws
\eqn\mazuruka{\eqalign{
&\bs B = -i\c,\cr
&\bbs B =i\bar\c,\cr
&\bs\bar\c = H -\fr{1}{2}[\w,B],\cr
&\bbs\c =H+\fr{1}{2}[\w,B],\cr
}\qquad
\eqalign{
&\bs\c = 0,\cr
&\bbs\bar\c = 0,\cr
&\bs H =-\fr{i}{2}[\w,\c],\cr
&\bbs H =-\fr{i}{2}[\w,\bar\c].\cr
}}
The ghost numbers of the various fields are given by
\eqn\zzh{\matrix{
&\hbox{Fields}
&A^\pr&A^\ppr&\p&\bar\p&\w&B&\c&\bar\c&H\cr
&\hbox{$U$ Number}  &0&0&1&1&2&-2&-1&-1&0\cr
&\hbox{$R$ Number}  &0&0&1&-1&0&0&1&-1&0\cr
}\;.
}

The action of $N=2$ TYM theory can be written in the form\foot{
The action of $N=1$ TYM theory can be written as $S=-i\dw W$
\WittenA\BS. The relation between $N=1$ and $N=2$ theories
can be most conveniently understood with an analogy to the exterior
derivative $d = \rd + \bar\rd$. An exact real $(p,p)$-form $\a=d\b$ on a
compact
K\"{a}hler manifold can be written as
$$
\a = \Fr{d d^c}{2}\g =\Fr{i\rd\bar\rd -i\bar\rd\rd}{2}\g =i\rd\bar\rd \g
$$
for some $(p-1,p-1)$-form $\r$, where $d^c \equiv -J^{-1}d J =i(\bar\rd
-\rd)$.}
\eqn\zzi{\eqalign{
S_{old}
= \Fr{\bs\bbs-\bs\bbs}{2} \CB_{\bf T}
=\Fr{\bs\bbs-\bbs\bs}{2}\left(-\Fr{1}{h^2}\int_X \tr B\wedge * F
-\Fr{1}{h^2}\int_X \tr \c\wedge * \bar\c
\right).
}}
Note that  $V$ has $(U,R)=(-2,0)$, so that the action
has $(U,R)=(0,0)$.
We find that
\eqn\gorby{\eqalign{
S_{old} =
&\Fr{1}{h^2}\!\int_X\!\!\! \tr\biggl[
  -H^{2,0}\wedge*( H^{0,2} +iF^{0,2})
  - H^{0,2}\wedge*( H^{2,0} +iF^{2,0})
  +i\c^{2,0}\wedge *\Dpp\bar\p
  \phantom{\biggl]}
  \cr
& +i\bar\c^{0,2}\wedge *\Dp\p
  + i[\w,\c^{2,0}]\wedge *\bar\c^{0,2}
  + i[\w,\c^{0,2}]\wedge *\bar\c^{2,0}
  -\fr{i}{2}B^{2,0}\wedge *\Dpp\Dpp\w
  \cr
& +\fr{i}{2}B^{0,2}\wedge *\Dp\Dp \w
  +\fr{1}{2}[\w,B^{2,0}]\wedge *[\w,B^{0,2}]
  -\biggl( 2H^{0}(H^0 +if)
  \cr
& -i\bar\c^{0}\L \Dpp\p
  -i\c^{0}\L \Dp\bar\p
  -2 i[\w,\c^0]\bar\c^0
  -\fr{1}{2}[\w,B^0][\w,B^0]
  \phantom{\biggl]}
  \cr
& +\ha B^{0}\L\left((i\Dp \Dpp - i\Dpp\Dp)\w -2[\p,\bar\p]\right)
  \biggr)\Fr{\o^2}{2!}\biggr]\;,
  \cr
}}
where $f =\ha\L F$ and $\L$ is adjoint to the wedge multiplication
of $\o$.

For the details how $N=2$ TYM theory (or TYM theory in general)
realizes the Donaldson invariants, we refer the  reader to
\ParkA\ (\WittenA\BS).
We will show in the subsequent subsection
that the old $N=2$ TYM theory realizes the algebraic part of
Donaldson's polynomals only.

\subsec{Problem of the on-shell invarinats}

An observable of $N=2$ supersymmetric TYM theory should be gauge
invariant as well as invariant under $\bs$ and $\bbs$.
The candidates of the nontrivial topological observables depending
on $H^2(X,\BZ)$ are
\eqn\moonlight{
\eqalign{
\tilde\o^{2,0}&=\Fr{1}{8\pi^2}\int_X\tr(\p\wedge\p)\wedge \o^{0,2}\;,\cr
\tilde\o^{0,2}&=\Fr{1}{8\pi^2}\int_X\tr(\bar\p\wedge\bar\p)\wedge \o^{2,0}\;,
\cr
\tilde\o^{1,1}&=
        \Fr{1}{4\pi^2}\int_X\tr(i\w F^{1,1}+\p\wedge\bar\p)\wedge\o^{1,1}\;,
        \cr
}}
where  $\o^{p,q} \in H^{p,q}(X,\BZ)$ and we generally denote
$\tilde\o^{r,s}$  as an $(r,s)$-form on $\CA$ (or of degree $(r,s)$).
Note that the above quantities are the components of the decompositions
of $\tilde\o^{(2)} \in H^2_\CG(\CA)$ and $\tilde\o^{(r,s)} \in
\O^{r,s}_\CG(\CA)$ with $r+s =2$.

As already noted in \ParkA, the only quantity which is
both $\bs$ and $\bbs$ invariant is $\tilde\o^{1,1}$.
The quantity
$\tilde\o^{0,2}$ is invariant only under $\bbs$ transformation while
$\tilde\o^{2,0}$ is invariant only under $\bs$ transformation, i.e.~
\eqn\yab{
\tilde\o^{1,1} \in H^{1,1}_\CG(\CA),\qquad
\tilde\o^{0,2} \in H^{0,2}_\CG(\CA),\qquad
\tilde\o^{2,0} \notin H^{2,0}_\CG(\CA).
}
The part $\Fr{1}{4\pi^2}\int_X \tr(\p\wedge\bar\p)\wedge\o^{1,1}$
of $\tilde\o^{1,1}$ is  a closed form on $\CA$.
Then, $\tilde\o^{1,1}$ is the equivariantly closed extension.
One the other hand, such a equivariant
extension of $\tilde\o^{2,0}$ is not possible since $\w$, which
is the generator of $\hbox{Fun}(Lie(\CG)$, is of degree $(1,1)$.
This is an example of the failure of the Hodge decompositions
in the equivariant sense.

The TYM theory realizes the Donaldson invariants by expectation
values of topological observables \WittenA. In the $N=2$ theory,
the quantities $\tilde\o^{2,0}$ and $\tilde\o^{0,2}$ are not in the
set of observables.

However, it is important to note that
$\tilde\o^{2,0}$ and $\tilde\o^{0,2}$ are nontrivial $\bs$ and $\bbs$
invariants if they  are restricted to the moduli space $\CM$
of ASD connections. The K\"{a}hler structure on $\CM$ garantees
that an $\bs$-invariant quantity is $\bbs$-invariant and vice versa.
{\it Put it differently, not all the elements of $H^{*,*}(\CM)$
can be obtained  from the elements of $H^{*,*}_\CG(\CA)$
by the restriction and the reduction.}
On the other hand,
the Donaldson invariants are cup products of (ordinary) cohomology
classes on $\CM$ evaluated on the fundamental homology cycle of $\CM$
provided with a suitable compactification.
And, the path integral of TYM theory is localized to
$\CM$. Thus, we should include $\tilde\o^{2,0}$ and $\tilde\o^{0,2}$
to realize the full invariants. Once the localization to $\CM$ and
a suitable procedure of including $\tilde\o^{2,0}$ and $\tilde\o^{0,2}$
are understood, it is sufficient to consider the $\bbs$ symmetry (that is,
the equivariant Dolbeault cohomology $H^{*,*}_\CG(\CA)$) only, due to the
familar Hodge decomposition theorem. These are the geometrical reasoning
underlying
the key procedure of Witten's breaking the $N=2$ supersymmetry down to
$N=1$ symmetry by introducing suitable mass terms \WittenB.

\subsec{The non-Abelian localization}

The $N=2$ HYM theory is another model for the Donaldson invariants
on a K\"{a}hler surface \ParkB, adopting Witten's non-Abelian equivariant
localization theorem \WittenC. It shares the same $N=2$ supersymmetry
(or the same structure of the Dolbeault equivariant cohomology)
with the $N=2$ TYM theory.
A version of the theorem states (in a field
theoretical context) that a path integral with an action functional given
by the normed square of the equivariant moment map of field configuration
can be expressed as sums of contributions of the critical points.
Such a path integral can be used to obtain cohomology rings of the reduced
phase space\foot{
This is an fastly growing subjects and we will not go into details.
Recent developements can be found in \Wu\JK\PW\KalkC\Verg.}.
The relavant path integrals are the partition function and the
expectation value of observables which correspond to equivariantly closed
form on the field configuration.

There would be two models of the equivariant localization, the original
de Rham model of Witten and the Dolbeault model. The $N=2$ HYM theory
is an example of the latter one. In terms of the de Rham model of
the equivariant localization, the entire (de Rham) cohomology rings on
the reduced phase space  can be obtained in principle. On the other
hand, a new problem arises in the Dolbeault model since not all the
Dolbeault cohomology classes on the reduced phase space would be obtained
from the elments of the Dolbeault equivariant cohomology.

the $N=2$ HYM theory on the K\"{a}hler surface\foot{On the other
hand, the $N=2$ HYM theory on a Riemann surface has no such a problem,
thus not really different with the original theory \WittenC,
since every $\bbs$ closed observables are $\bs$ closed \ParkA.}
One of the main purposes of this paper is to eliminate that
problem  in the $N=2$ HYM theory on the K\"{a}hler surface\foot{On the other
hand, the $N=2$ HYM theory on a Riemann surface has no such a problem,
thus not really different with the original theory \WittenC,
since every $\bbs$ closed observables are $\bs$ closed \ParkA.}.
Clearly, this is closely related to the similar problem
of the $N=2$ TYM theory.

\newsec{New Construction}

In this section, we construct a new $N=2$ TYM theory to overcome
the problem of the on-shell invariants.
In the new construction, we will impose different transformation
laws for anti-ghost multiplets.

\subsec{Action}

We introduce a commuting anti-ghost
$B^0\in \O^0(\gE)$
in the adjoint representation with
$(U,R) = (-2,0)$.
Then \kkg\ leads us to
multiplet $(B^0,i\c^0, -i\bar\c^0, H^0)$ with
transformation laws
\eqn\maz{\eqalign{
&\bs B^0 = -i\c^0,\cr
&\bbs B^0 =i\bar\c^0,\cr
&\bs\bar\c^0 = H^0 -\fr{1}{2}[\w,B^0],\cr
&\bbs\c^0 =H^0+\fr{1}{2}[\w,B^0],\cr
}\qquad
\eqalign{
&\bs\c^0 = 0,\cr
&\bbs\bar\c^0 = 0,\cr
&\bs H^0 =-\fr{i}{2}[\w,\c^0],\cr
&\bbs H^0 =-\fr{i}{2}[\w,\bar\c^0].\cr
}}
We also introduce an anti-commuting anti-ghost $\c^{2,0} \in
\O^{2,0}(\gE)$ with $(U,R) = (-1,1)$ and an anti-commuting
anti-ghost $\bar\c^{0,2} \in \O^{0,2}(\gE)$ with $(U,R) = (-1,-1)$
with transformation laws
\eqn\xxx{\eqalign{
&\bs\c^{2,0} = 0,\cr
&\bbs\c^{2,0} = H^{2,0},\cr
&\bs\bar\c^{0,2}=H^{0,2},\cr
&\bbs\bar\c^{0,2}=0,\cr
}\qquad
\eqalign{
&\bs H^{2,0} = -i[\w,\c^{2,0}],\cr
&\bbs H^{2,0} =0,\cr
&\bs H^{0,2} = 0,\cr
&\bbs H^{0,2} =- i[\w,\bar\c^{0,2}].\cr
}}
One can easily check that these satisfy the commutation relations
\kkg.

%\lin{Action Functional}

Now, the most general form of new $N=2$ supersymmetric action is
\eqn\neww{
S =  i\bs\, \overline{V} + i\bbs\, V + \Fr{(\bs\bbs -\bbs\bs)}{2} \CB,
}
where $V$ and $\overline{V}$ should be $\bbs$ and $\bs$ closed quantities
 with  $(U,R)$ numbers $(-1,1)$ and $(-1,-1)$, respectively.
One finds the following unique choices
\eqn\new{
\eqalign{
&\overline{V} =-\Fr{1}{h^2}\int_X \tr \bar\c^{0,2}\wedge * F^{2,0},\cr
&V =-\Fr{1}{h^2}\int_X \tr \c^{2,0}\wedge * F^{0,2},\cr
&\CB = -\Fr{1}{h^2}\int_X\tr\left( B^0 f +\a \c^0\bar\c^0 \right)\o^2
-\Fr{2\b}{h^2}\int_X\tr \c^{2,0}\wedge *\bar\c^{0,2},\cr
}}
where $\a,\b=0,or\; 1$.
For $\a=\b=1$, we find
\eqn\gorby{\eqalign{
S =
&\Fr{1}{h^2}\!\int_X\!\!\! \tr\biggl[
  -2H^{2,0}\wedge* H^{0,2}
  -iH^{2,0}\wedge *F^{0,2}
  -iH^{0,2}\wedge* F^{2,0}
  + 2i[\w,\c^{2,0}]\wedge *\bar\c^{0,2}
  \phantom{\biggl]}
  \cr
& +i\c^{2,0}\wedge *\Dpp\bar\p
  +i\bar\c^{0,2}\wedge *\Dp\p
  -\biggl( 2H^{0} H^0 +2iH^0f
  -2 i[\w,\c^0]\bar\c^0
  \cr
& +\ha B^{0}\L\left((i\Dp \Dpp - i\Dpp\Dp)\w -2[\p,\bar\p]\right)
  -\fr{1}{2}[\w,B^0][\w,B^0]
  \phantom{\biggl]}
  \cr
& -i\bar\c^{0}\L \Dpp\p
  -i\c^{0}\L \Dp\bar\p
  \biggr)\Fr{\o^2}{2!}\biggr]\;.
  \cr
}}
We can integrate out $H^{2,0}$, $H^{0,2}$ and $H^{0}$ from the
action by setting $H^{2,0} =-\fr{i}{2} F^{2,0}$,
$H^{0,2} =-\fr{i}{2} F^{0,2}$ and $H^0 = -\fr{i}{2}f^0$,
or by the Gaussian integral, which
leads to modified transformation laws
\eqn\jane{
\eqalign{
&\bs\bar\c^{0,2} = -\fr{i}{2} F^{0,2}\;,\cr
&\bs\bar\c^{0} = -\fr{i}{2} f-\ha[\w,B^{0}]\;,\cr
}\qquad\eqalign{
&\bbs\c^{2,0} = -\fr{i}{2} F^{2,0}\;,\cr
&\bbs\c^{0} = -\fr{i}{2} f+\ha[\w,B^{0}] \;.\cr
}}
One can see that the locus of $\bs$ and $\bbs$ fixed points in
the above transformations is precisely the space of ASD connections.
Now we can rewrite the action as
\eqn\gad{\eqalign{
S =
&\Fr{1}{h^2}\!\int_X\!\!\! \tr\biggl[
  -\ha F^{2,0}\wedge* F^{0,2}
  +i\c^{2,0}\wedge *\Dpp\bar\p
  +i\bar\c^{0,2}\wedge *\Dp\p
  - 2i[\w,\c^{2,0}]\wedge *\bar\c^{0,2}
  \phantom{\biggl]}
  \cr
& -\biggl( \fr{1}{2}f^2
  -2i[\w,\c^0]\bar\c^0
  -i\bar\c^{0}\L \Dpp\p
  -i\c^{0}\L \Dp\bar\p
  -\fr{1}{2}[\w,B^0][\w,B^0]
  \cr
& +\ha B^{0}\L\left((i\Dp \Dpp - i\Dpp\Dp)\w -2[\p,\bar\p]\right)
  \biggr)\Fr{\o^2}{2!}\biggr]\;.
  \cr
}}

One can easily check that this new theory shares almost
all the properties with the old theory studied in \ParkA. A notable
difference between the two theories is that $\c^{2,0}$ ($\bar\c^{0,2}$)
is no  longer $\bs$-exact ($\bbs$-exact) in the new setting.

\rem{Remark}

One may wonder why the transformation laws for the antighosts multiplets
Eq.(3.1) and Eq.(3.2) are different. To understand this, we should
recall the Atiyah-Jeffrey's intepretation of TYM theory \AJ\ based on
the Mathai-Quillen formalism \MQ. Consider an
infinite dimensional vector bundle $\CQ$ over $\CA/\CG$
whose section $s$ is $s(A) = - F^+(A)$ where $F^+$ is the self-dual
part of the curvature. The moduli space $\CM$ of ASD connections
is the zero-locus of the section $s$.
In our case, we can
decompose the section $s$ (the bundle $\CQ$) according to the
decomposistions
$
F^+(A) = F^{2,0}(A^\pr)\oplus f(A^\pr, A^\ppr)\o \oplus F^{0,2}(A^\ppr)
$.
Roughly speaking, the anti-ghosts
live in the dual space of the fiber $V$ of $\CQ$ \Kalk.
We have introduced
the commuting anti-ghost $B^0$  for the constraint
$f(A^\pr,A^\ppr)=\Fr{1}{2}\L F^{1,1}(A^\pr,A^\ppr) =0$
and the anti-commuting anti-ghosts $\bar\c^{0,2}$
and $\c^{2,0}$ for the constraints $F^{2,0}(A^\pr) = 0$ and
$F^{0,2}(A^\ppr) = 0$, respectively.
The underlying reason for the different transformation
laws Eq.(3.1) and Eq.(3.2) is that
that $F^{2,0}(A^\pr)$ and $F^{0,2}(A^\ppr)$
depend only on $A^\pr$ and on $A^\ppr$, respectively,
while $f(A^\pr,A^\ppr)$ depends both on $A^\pr$ and $A^\ppr$.
The detail will appear elsewhere \HPb.

\subsec{Fermionic zero modes}

Important properties common to both old and new $N=2$ topological
Yang-Mills theories are the roles of fermionic zero-modes.
We will briefly recall the results of \ParkA.
The related mathematical topics can be found in \DK\Kobayashi.

It is convenient to use
the language of holomorphic vector bundles. It is well
known that an ASD connection $A$ endows $E$ with a holomorphic structure
$\CE_A$ of given topological type.
Let $\hbox{End}_0(\CE_A)$ be the trace-free endomorphism bundle
of $\CE_A$. It turns out that  zero-modes of $\bar\c_0$, $\bar\p$
and $\bar\c^{0,2}$ define elements of $H^{0}(\hbox{End}_0(\CE_A))$,
$H^{1}(\hbox{End}_0(\CE_A))$ and $H^{2}(\hbox{End}_0(\CE_A))$,
respectively. The formal complex dimension of the moduli space
$\CM$ is $(-{\bf h}^{0,0} +{\bf h}^{0,1} -{\bf h}^{0,2})$,
where ${\bf h}^{0,p} = \hbox{dim}_\BC H^{p}(\hbox{End}_0(\CE_A))$.

Since the fermionic zero-modes of $(\bar\c_0,\bar\p,\bar\c^{0,2})$
carry the $U$-charge $(-1,1,-1)$, the half of the net violation
$\D U/2$ of the $U$-number in the path integral measure is equal to
the formal complex dimension.  It is important to note that there
is no net $R$ number violation in the path integral measure \ParkA.
We assume, throughout this paper, that there exist the zero-modes pairs
of $\p$ and $\bar\p$ only. Then the moduli space is a smooth K\"{a}hler
manifold with complex dimension $d = 4k -3(1 +p_g)$,
identical to the number of $\bar\p$ zero-modes.

It is convenient to introduce quantum
operators $\hat U$ and $\hat R$ such that
\eqn\caccc{
\eqalign{
&\hat U \c^{0} = u^{-1}\c^{0},\cr
&\hat U \bar\c^{0} = u^{-1}\bar\c^{0},\cr
&\hat R \c^{0} = r\c^{0}, \cr
&\hat R \bar\c^{0} = r^{-1}\bar\c^{0},\cr
}\qquad
\eqalign{
&\hat U \p = u\p,\cr
&\hat U \bar\p = u\bar\p,\cr
&\hat R \p = r\p, \cr
&\hat R \bar\p = r^{-1}\bar\p,\cr
}\qquad
\eqalign{
&\hat U \c^{2,0} = u^{-1}\c^{2,0},\cr
&\hat U \bar\c^{0,2} = u^{-1}\bar\c^{0,2},\cr
&\hat R \c^{2,0} = r\c^{2,0}, \cr
&\hat R \bar\c^{0,2} = r^{-1}\bar\c^{0,2},\cr
}\qquad
\eqalign{
&\hat U B^{0} = u^{-2}R^{0},\cr
&\hat R B^{0} = B^{0},\cr
&\hat U \w = u^{2}\w^{0}, \cr
&\hat R \w =    \w.\cr
}
}

Then the action $S$ is invariant under the transformations generated by
$\hat U$ and $\hat R$.
Now the fermionic part $\CD X_f$ of path integral measure, after integrating
out every non-zero modes, reduces to
\eqn\cadx{
\CD\hat X_f =\prod_{i}^{d} \p_i\bar\p_i,
}
which transforms, under $\hat U$ and $\hat R$, as
\eqn\cae{
\CD\hat X_f \rightarrow \CD\hat X_f u^{-2d}.
}
Thus, the expectation value of topological observables
\eqn\caf{
\left<\prod_{i}^{n} \tilde\o^{r_i,s_i} \right>
=\Fr{1}{\hbox{vol}(\CG)}\int \CD X e^{-S}\cdot \prod_{i=1}^{r}
\tilde\o^{r_i,s_i},
}
evaluated with the action $S$ vanishes unless (see \WittenB\ for
related analysis)
\eqn\cag{
\sum_{i=1}^{n}(r_i + s_i) = 2d\;\hbox{ and }\; \sum_{i=1}^{n}(r_i-s_i)
\quad \Rightarrow\quad \sum_{i=1}^{r}(r_i,s_i) = (d,d).
}
This selection rule is, more or less, identical to the statement that
the Donaldson invariants are pure Hodge type of $(d,d)$ \BrA\Tuyrin.

\subsec{Including the on-shell observables}

There is a nice method to deal with on-shell
invariant quantities (pp $149-151$ of \WittenD).
To use
$\tilde\o^{0,2}$ and $\tilde\o^{2,0}$,
we should change the transformation laws \xxx\ as
\eqn\caa{
\bbs\bar\c^{0,2} = \Fr{h^2}{4\pi^2}\w\o^{0,2},\qquad
\bs\c^{2,0} = \Fr{h^2}{4\pi^2}\w\o^{2,0}.
}
and add the terms
\eqn\cab{
-\Fr{1}{8\pi^2}\int_X\tr(\p\wedge\p)\wedge \o^{0,2}
-\Fr{1}{8\pi^2}\int_X\tr(\bar\p\wedge\bar\p)\wedge \o^{2,0},
}
to the action \gad. Then the action is  both $\bs$ and $\bbs$
invariant with the modified transformation laws of \caa.

However, we can not use the above prescription in the old construction.
Since $\w$ is both $\bs$ and $\bbs$ closed,
we have $\bbs^2\bar\c^{0,2}= \bs^2\c^{2,0} = 0$.
However, Eqs.\mazuruka\ and \caa\ shows that
$\bbs^2 B^{2,0} = i\bbs\bar\c^{2,0}\neq 0$.
Thus, the changes of the transformation laws as \caa\
do violate the relations $\bs^2=\bbs^2 = 0$.
This is why the old theory realizes only the algebraic part
of  Donaldson's polynomials, defined by algebraic cycles
which are Poincar\'{e} dual to  elements of $H^{1,1}(S,\BZ)$.

At this point, it is sufficient to consider only $N=1$ part of
the supersymmetry as explained in Sect.$2.2$.
We choose $\bbs$ symmetry. Since $\tilde\o^{0,2}$ is $\bbs$
invariant and $\tilde\o^{2,0}$ is $\bbs$ invariant modulo $\bar\c^{0,2}$
equation of motion, it is sufficient to change the transformation law
for $\bar\c^{0,2}$ only in Eq.(3.2) as
\eqn\cac{
\bbs\bar\c^{0,2} = \Fr{h^2}{4\pi^2}\w\o^{0,2},
}
and add
\eqn\cad{
-\Fr{1}{8\pi^2}\int_X\tr(\p\wedge\p)\wedge \o^{0,2} \equiv -\tilde\o^{2,0},
}
to the action \gad;
\eqn\gadp{\eqalign{
S^\pr =
&\Fr{1}{h^2}\!\int_X\!\!\! \tr\biggl[
  -\ha F^{2,0}\wedge* F^{0,2}
  +i\c^{2,0}\wedge *\Dpp\bar\p
  +i\bar\c^{0,2}\wedge *\Dp\p
  - 2i[\w,\c^{2,0}]\wedge *\bar\c^{0,2}
  \phantom{\biggl]}
  \cr
& -\biggl( \fr{1}{2}f^2
  -2i[\w,\c^0]\bar\c^0
  -i\bar\c^{0}\L \Dpp\p
  -i\c^{0}\L \Dp\bar\p
  -\fr{1}{2}[\w,B^0][\w,B^0]
  \cr
& +\ha B^{0}\L\left((i\Dp \Dpp - i\Dpp\Dp)\w -2[\p,\bar\p]\right)
  \biggr)\Fr{\o^2}{2!}\biggr]
  -\Fr{1}{8\pi^2}\int_X\tr(\p\wedge\p)\wedge
  \o^{0,2}
  .\cr
}}
Note that this action has actually the full $N=2$ symmetry.
Clearly, the action $S^\pr=S -\tilde\o^{2,0}$ is not invariant under the
transformations generated by $\hat U$ and $\hat R$. However, the path
integral measure, after integrating out every non-zero-modes, is identical
to the one defined by the action $S$, since the additional term does not
change the equations of zero-modes.
Therefore the partition function $<1>^\pr$ for the action $S^\pr$
can be interpreted as the following expectation value;
\eqn\cah{
<1>^\pr  = \left<\sum_{n=0}^{\infty} \Fr{1}{n!}(\tilde\o^{2,0})^n \right>,
}
evaluated in the theory with the action $S$. Clearly, this is non-zero only
for $d=0$ and identical to $<1>$.

One can further add the following term to $S^\pr$ maintaining the
$\bbs$ symmetry,
\eqn\wga{
\bbs\biggl( \int_X \tr(B^0 \bar\c^{0,2})\wedge\o^{2,0}\biggr)
=
i\int_X\tr(\bar\c^0\bar\c^{0,2})\wedge\o^{2,0}
 +\Fr{h^2}{4\pi^2}\int_X \tr(B^0\w)\o^{2,0}\wedge\o^{0,2}.
}
Adding these terms will explicitly break the $N=2$ supersymmetry down to the
$N=1$  supersymmetry (the $\bbs$-symmetry).
The new $\bbs$-invariant action is
\eqn\yad{
S^\ppr = S
-\Fr{1}{8\pi^2}\int_X\tr(\p\wedge\p)\wedge\o^{0,2} +
i\int_X\tr(\bar\c^0\bar\c^{0,2})\wedge\o^{2,0}
 +\Fr{h^2}{4\pi^2}\int_X \tr(B^0\w)\o^{2,0}\wedge\o^{0,2}.
}

The above procedures to obtain $S^\pr$ and $S^\ppr$ from
the original $N=2$ supersymmetric action was directly
motivated by Sect.~3 of Ref.\WittenB. Adding $-\tilde\o^{2,0}$
to the action $S$ gives the bare mass to $\p$. Adding \wga\
to $S^\pr$ by breaking the $N=2$ symmetry down to $N=1$ induces
the mass gap to $\w$. This can be most easily seen by the $B^0$
equation of motion for the action $S^\ppr$,
\eqn\yae{
(\Da^*\Da + \Fr{h^4}{\pi^2}m\bar m)\w + 2\L([\p,\bar\p]) =0,
\rightarrow
<\w> = \Fr{-2\L([\p,\bar\p ])}{\Da^*\Da + \Fr{h^4}{\pi^2}m\bar m},
}
where we have used the K\"{a}hler identities
\eqn\kaehler{
\Dpp^* = i[\Dp,\L], \qquad \Dp^* = -i[\Dpp,\L],
}
and set $\o^{2,0}\wedge\o^{0,2}= m\bar m (\o\wedge\o)$.
The mass gap of the theory was crucial in Witten's calculation
in \WittenB.
Of course, the mass gap disappear in the vanishing locus of $\o^{0,2}$.

\newsec{Deformations to Holomorphic Yang-Mills Theories}

We now turn to HYM theory.
Since the terms which are proportional to the K\"ahler form
are identical in  the old and new actions $S_{old}$ and $S$,
we can repeat the procedure in \ParkB\ to obtain $N=2$ HYM theory.
It is convenient to choose delta function gauge by setting $\a=\b=0$
in \new. Now the action for $N=2$ TYM theory  is
\eqn\hba{\eqalign{
S =
&\Fr{1}{h^2}\!\int_X\!\!\! \tr\biggl[
  -iH^{2,0}\wedge *F^{0,2}
  -iH^{0,2}\wedge* F^{2,0}
  +i\c^{2,0}\wedge *\Dpp\bar\p
  +i\bar\c^{0,2}\wedge *\Dp\p
  \cr
& -\biggl(2iH^0f
  -i\bar\c^{0}\L \Dpp\p
  -i\c^{0}\L \Dp\bar\p
  +\ha B^{0}\L\left((i\Dp \Dpp - i\Dpp\Dp)\w -2[\p,\bar\p]\right)
  \biggr)\Fr{\o^2}{2!}\biggr]\;.
  \cr
}}
Then, the action of $N=2$ HYM theory becomes
\eqn\haa{\eqalign{
S_H =
&\Fr{1}{h^2}\!\int_X\!\!\! \tr\biggl[
  -iH^{2,0}\wedge *F^{0,2}
  -iH^{0,2}\wedge* F^{2,0}
  +i\c^{2,0}\wedge *\Dpp\bar\p
  +i\bar\c^{0,2}\wedge *\Dp\p
 \biggl]
  \cr
&-\Fr{1}{4\pi^2}\int_X\tr(i\w F +\p\wedge\bar\p)\wedge\o
 -\Fr{\e}{8\pi^2}\int_X\tr(\w^2)\Fr{\o^2}{2!}.
  \cr
}}
This is equivalent to the action studied in \ParkB.
The difference is that $\c^{2,0}$ and $\bar\c^{0,2}$
are no longer BRST exact in this new setting.

Since $N=2$ HYM theory has the same $N=2$ supersymmetry and the
same topological observables  as those of $N=2$ TYM theory, we can repeat
the same procedure to deal with the on-shell invariant quantities.
It is sufficient to consider the $N=1$ part of the symmetry and
we, once again, consider the $\bbs$ symmetry.
Adding \cad\ to the action $S_H$, we have a new action
\eqn\hab{\eqalign{
S^\pr_H =
&\Fr{1}{h^2}\!\int_X\!\!\! \tr\biggl[
  -iH^{2,0}\wedge *F^{0,2}
  -iH^{0,2}\wedge* F^{2,0}
  +i\c^{2,0}\wedge *\Dpp\bar\p
  +i\bar\c^{0,2}\wedge *\Dp\p
 \biggl]
  \cr
&-\Fr{1}{4\pi^2}\int_X\tr(i\w F +\p\wedge\bar\p)\wedge\o
 -\Fr{\e}{8\pi^2}\int_X\tr(\w^2)\Fr{\o^2}{2!}
  \cr
&-\Fr{1}{8\pi^2}\int_X\tr(\p\wedge\p)\wedge \o^{0,2},
}}
where the change of the transformation of $\bbs$ as \cac\ is
understood. Of course, we start from the action $S^\pr$ (in the delta
function gauge) and then define the mapping to the HYM theory.
Both procedures give the identical result.

The partition function $Z(\e)_d$ of the HYM theory with action $S_H$
is a generating functional
\eqn\hac{
Z(\e)_d =
\sum_{r,s}^{r+2s =d}\Fr{\e^s}{r! s!}\left<\tilde\o^r\Theta^s\right>
+\;\hbox{exponentially small terms},
}
where
\eqn\had{\eqalign{
&\tilde\o =\Fr{1}{4\pi^2}\int_X\tr(i\w F +\p\wedge\bar\p)\wedge\o,\cr
&\Theta = \Fr{1}{8\pi^2}\int_X\tr(\w^2)\Fr{\o^2}{2!},
}}
and the exponentially small terms are the contributions of the higher
critical points \WittenC\ParkB.
Note that  the partition function $Z^\pr(\e)_d$
with action $S^\pr_H$ is identical to $Z(\e)_d$;
\eqn\hae{\eqalign{
Z^\pr(\e)_d = \left<\sum_{n=0}^{\infty} \Fr{1}{n!}(\tilde\o^{2,0})^n
\right>_{H}
&=\sum_{n=0}^{\infty} \Fr{1}{n!} \sum_{r,s}^{r+2s =d-n}\Fr{\e^s}{r! s!}
\left<(\tilde\o^{2,0})^n\tilde\o^r\Theta^s\right> + ...\cr
&=\sum_{r,s}^{r+2s =d}\Fr{\e^s}{r! s!}
\left<\tilde\o^r\Theta^s\right> + ...\cr
&=  Z(\e)_d.
}
}

However, the HYM theories with the actions $S_H$ and $S^\pr_H$ have
different localizations. The $H^{2,0}$, $H^{0,2}$, $\c^{2,0}$ and
$\bar\c^{0,2}$ integrations localize the both theories to
$T\CA^{1,1}$,
\eqn\yaf{
F^{2,0} = F^{0,2} = \Dp\p = \Dpp\bar\p =0.
}
The $\w$ equations of the motion for the both
actions $S_H$ and $S^\pr_H$  give
\eqn\yag{
2if + \e\w = 0.
}
Then, a fixed point equation of the basic supersymmetry leads to
\eqn\yah{
\bbs\p = -i \Dp \w =0
\Longrightarrow
\Dp f = 0
\Longleftrightarrow
\Da f = 0.
}
However, the theory with the action $S^\pr_H$ has an additional
fixed point equation
due to Eq.\cac
\eqn\yai{
\bbs\bar\c^{0,2} = \Fr{h^2}{4\pi^2}\w\o^{0,2} = 0.
}
This shows that in the vanishing locus $C \subset X$ of $\o^{0,2}$ the same
localization governs the two theories, while in the complement of
$C$ the theory with the action $S^\pr_H$  is localized to the instanton
$(f =0)$. Due to the supersymmetry, we also have
\eqn\yaj{
f = 0 \Longrightarrow  \bbs f = \Dpp^*\bar\p = 0.
}

Finally, we note that the HYM theory with action $S^\pr_H$ is
entirely equivalent to the TYM theory with ation $S^\pr$ for
hyper-K\"{a}hler surfaces. There will be no
contributions of
higher critical points, since those manifolds have only one holomorphic
harmonic two-form which is nowhere vanishing. This may be related
to a general fact that the twisting of $N=2$ supersymmetric theory
does not change anything on a manifold with trivial canonical line bundle
\WittenB\WittenD.

\subsec{Deformation from the action $S^\ppr$.}

One can also start with the TYM theory with action $S^\ppr$ (in the
delta function gauge) which has  the $\bbs$ symmetry only.
We will show that there is a suitable deformation of $S^\ppr$ to
the HYM theory with action $S^\pr_H$.

We can add to the action $S^\ppr$
an $\bbs$-exact term  maintaining the $\bbs$ symmetry,
\eqn\yak{
i\bbs\biggl(-\Fr{4t}{h^2}\int_X\Fr{\o^2}{2!}\tr(B^0\c^0)\biggr)
=-\Fr{4t}{h^2}\int_X\Fr{\o^2}{2!}\tr(\c^0\bar\c^0 +iH^0B^0),
}
which leads to a family of $\bbs$ symmetric action $S^\ppr(t)$
\eqn\yal{\eqalign{
S^\ppr(t)
=
&S^\ppr + i\bbs\biggl(-\Fr{4t}{h^2}\int_X\Fr{\o^2}{2!}\tr(B^0\c^0)\biggr)\cr
=
& \Fr{1}{h^2}\!\int_X\!\!\! \tr\biggl[
  -iH^{2,0}\wedge *F^{0,2}
  -iH^{0,2}\wedge* F^{2,0}
  +i\c^{2,0}\wedge *\Dpp\bar\p
  +i\bar\c^{0,2}\wedge *\Dp\p
  \phantom{\biggl]}
  \cr
& -\biggl(2iH^0(f+2t B^0)
  +\Ha B^{0}\L\left((i\Dp \Dpp - i\Dpp\Dp)\w -2[\p,\bar\p]\right)
  +4t\c^0\bar\c^0
  \phantom{\biggl]}
  \cr
& -i\bar\c^{0}\L \Dpp\p
  -i\c^{0}\L \Dp\bar\p
  \biggr)\Fr{\o^2}{2!}\biggr]
  -\Fr{1}{8\pi^2}\int_X\tr(\p\wedge\p)\wedge\o^{0,2}
  \phantom{\biggl]}
  \cr
& +i\int_X\tr(\bar\c^0\bar\c^{0,2})\wedge\o^{2,0}
  +\Fr{h^2}{4\pi^2}\int_X \tr(B^0\w)\o^{2,0}\wedge\o^{0,2}.
  \cr
}
}
After integrating $B^0$, $\c^0$ and $\bar\c^0$ out, we have
\eqn\yam{\eqalign{
S^\ppr(t)
=
& \Fr{1}{h^2}\!\int_X\!\!\! \tr\biggl[
  -iH^{2,0}\wedge *F^{0,2}
  -iH^{0,2}\wedge* F^{2,0}
  +i\c^{2,0}\wedge *\Dpp\bar\p
  +i\bar\c^{0,2}\wedge *\Dp\p
  \phantom{\biggl]}
  \cr
& +\biggl(
  \Fr{1}{4t}f\Da^* \Da \w -\Fr{1}{2t}f\L([\p,\bar\p])
  +\Fr{1}{4t}(\Dp^*\p)(\Dpp^*\bar\p)
  \biggr)\Fr{\o^2}{2!}\biggr]
  -\Fr{1}{8\pi^2}\int_X\tr(\p\wedge\p)\wedge\o^{0,2}
  \cr
& -\Fr{1}{4 t}\int_X\tr(\Dpp^*\bar\p\, \bar\c^{0,2})\wedge\o^{2,0}
  -\Fr{h^2}{8\pi^2 t}\int_X \tr(\w f)\o^{2,0}\wedge\o^{0,2},
  \cr
}
}
where we have used the K\"{a}hler identities \kaehler.

Now we examine what kind of localization governs the deformed theory
with action $S^\ppr(t)$.
The $H^{2,0}$ and $H^{0,2}$ integration localize the theory
to $\CA^{1,1}$. The $\w$ integration gives
\eqn\yan{
-\Da^*\Da f \Fr{\o^2}{2} - \Fr{h^4}{2\pi^2}f\o^{2,0}\wedge\o^{0,2} = 0
\Longrightarrow
-\int_X \tr(\Da f\wedge * \Da f)
-\Fr{h^4}{\pi^2} \int_X m\bar m \tr(f*f) = 0.
}
Thus, the fixed points of the deformed theory are
$\Da f = 0$ at the vanishing locus $C$ of $\o^{0,2}$ and
instantons $(f =0)$ in the complement of $C$.
We see that the deformed theory has the same bosonic
fixed point with the HYM theory with the action $S^\pr_H$.
The $\c^{2,0}$ and $\bar\c^{0,2}$ integrations give
\eqn\yao{
i\Dpp\bar\p =0,\qquad
i\Dp\p + \Fr{h^2}{4t}\Dpp^*\bar\p\wedge\o^{2,0} =0.
}
In the locus $C$, the above equations reduce to
\eqn\yap{
\Dp\p =0,\qquad \Dpp\bar\p = 0,
}
while in the complement of $C$ we have additional equation
\eqn\yaq{
\Dpp^*\bar\p = 0.
}
This coincides to the bosonic fixed point $f=0$ in the complement
of $C$
\eqn\yar{
\bbs f = 0 \Longrightarrow \Dpp^*\bar\p = 0.
}
Thus,  the deformed theory has the same fixed points with
the HYM theory with action $S^\pr_H$.
Then, the final step of the deformation is to consider the expectation
value of the observable $\exp(\tilde\o + \e\Theta)$ with $t=\infty$
limit, which leads to the action $S^\pr_H$.

It is interesting to note that the action $S^\pr_H$ actually has the
full $N=2$ symmetry. During the deformation of the $N=1$ symmetric
TYM action $S^\ppr$ to the HYM theory, the broken $N=1$ symmetry
(the $\bs$-symmetry) is restored. We do not know whether this
has any physical application.

Anyway, it is sufficient to consider the $\bbs$-symmetry only. If we
want to maintain the full symmetry explicitly, we should change the
transformation laws as \caa\ and add \cab\ to the action $S_H$, which
leads to
\eqn\yas{
S^\ppr_H = S_H -\tilde\o^{2,0} -\tilde\o^{0,2}.
}
The partition function $Z^\ppr(\e)$ with the action $S^\ppr_H$ is
identical to
\eqn\yat{
Z^\ppr(\e)
= \left<\sum \Fr{1}{n!}(\tilde\o^{0,2})^n \right>^\pr_H
= \left<\sum \Fr{1}{n!n!}(\tilde\o^{0,2}\tilde\o^{0,2})^n \right>_H,
}
where $<..>^\pr_H$ denotes the expectation value evaluated with
the action $S^\pr_H$. In the above identification, considering
$\bbs$-symmetry only is understood such that $\tilde\o^{0,2}$
can be an observable.

\subsec{A simple calculation}

Now we determine the Donaldson polynomial invariants on
$H^{0,2}(X,\BZ)\oplus H^{2,0}(X,\BZ)$.

We consider the partition function $Z^\pr(\e)_d$ of HYM theory
with action $S^\pr_H$.
\eqn\haf{\eqalign{
Z^\pr(\e)_d =&
\Fr{1}{\hbox{vol}(\CG)}\int
\!\!\CD \! A^\pr\,\CD \! A^\ppr\,
\CD \p\,\CD \bar\p\,\CD \w\,\CD H^{2,0}\,\CD H^{0,2}\,
\CD\c^{2,0}\, \CD \bar\c^{0,2}
\cr
&\times\exp\biggl(
  \Fr{1}{h^2}\!\int_X\!\!\! \tr\biggl[
  iH^{2,0}\wedge *F^{0,2}
  +iH^{0,2}\wedge* F^{2,0}
  -i\c^{2,0}\wedge *\Dpp\bar\p
  -i\bar\c^{0,2}\wedge *\Dp\p
  \biggl]
  \cr
&
+\Fr{1}{4\pi^2}\int_X \tr\bigl(i\w F^{1,1}
+\p\wedge\bar\p \bigr)\wedge\o
+\Fr{\e}{8\pi^2}\int_X \Fr{\o^2}{2!}\tr\w^2
\phantom{\biggl]}\cr
&+\Fr{1}{8\pi^2}\int_X\tr(\p\wedge\p)\wedge\o^{0,2}
\biggr). \cr
}
}

It is more convenient to represent
$\tilde\o^{0,2}$ and $\tilde\o^{2,0}$
by
\eqn\saa{
\tilde\o^{0,2}= \Fr{1}{8\pi^2}\int_{\G} \tr(\bar\p\wedge\bar\p),
\qquad
\tilde\o^{2,0}= \Fr{1}{8\pi^2}\int_{\bar\G} \tr(\p\wedge\p),
}
where $\G$ and $\bar\G$ denote homology cycles Poincar\'{e} dual to
$\o^{2,0}$ and $\o^{0,2}$, respectively.
Now we want to determine the expectation value
$\left<(\tilde\o^{0,2})^m\right>^\pr_H$ evaluated
in the HYM theory with action $S^\pr_H$
\eqn\sab{\eqalign{
\left<(\tilde\o^{0,2})^m\right>^\pr_H
&=\sum_{n=0}^{\infty}\Fr{1}{n!}
\left<(\tilde\o^{0,2})^m(\tilde\o^{2,0})^n\right>_H
\cr
&=\sum_{n=0}^{\infty}\Fr{1}{n!}
\sum_{r,s}^{r+2s=d-n-m}\Fr{\e^s}{r!s!}
\left<(\tilde\o^{0,2})^m(\tilde\o^{2,0})^n
\tilde\o^r\Theta^s
\right> +\cdots
\cr
&=\Fr{1}{m!}
\sum_{r,s}^{r+2s=d-2m}\Fr{\e^s}{r!s!}
\left<(\tilde\o^{0,2}\tilde\o^{2,0})^m
\tilde\o^r\Theta^s
\right> +\cdots
\cr
&= \Fr{1}{m!}\left<(\tilde\o^{0,2}\tilde\o^{2,0})^m\right>_H.
}}
Thus, we consider
\eqn\sac{\eqalign{
\left<(\tilde\o^{0,2}\tilde\o^{2,0})^m\right>_H
=& \Fr{1}{\hbox{vol}(\CG)}\int
\!\!\CD \! A^\pr\,\CD \! A^\ppr\,
\CD \p\,\CD \bar\p\,\CD \w \cdots
\cr
&\times\exp\biggl(\cdots +
\Fr{1}{4\pi^2}\int_X \tr\bigl(i\w F^{1,1} +
\p\wedge\bar\p \bigr)\wedge\o
+\Fr{\e}{8\pi^2}\int_X \Fr{\o^2}{2!}\tr\w^2
\biggr)\cr
&\times
\left(
\Fr{1}{8\pi^2}\int_{\bar\G}\tr\p\wedge\p
\cdot\Fr{1}{8\pi^2}\int_{\G}\tr\bar\p\wedge\bar\p
\right)^m.
}}

We note that $\p$ and $\bar\p$ are coupled  as free fields with
the trivial propagator,
\eqn\sad{
<\p^a_i(x)\bar\p^b_{\bar j}(y)> = -i4\pi^2\e_{i\bar{j}}\d^{ab}\d^4(x-y).
}
To be more precise, this amounts to perform gaussian integrals
in the action $S^\ppr_H$.
The $\p$ and $\bar\p$ are obviously coupled as free field for the vanishing
locus $C$ of $\o^{0,2}$ in $X$. The actual calculation (using the K\"{a}hler
identites) shows that they are coupled as free field even in the
complement of $C$ if $\Dpp^*\bar\p = \Dp^*\p = 0$ which are garanteed, as
explained in previous subsection.
Upon performing the $\p$ and $\bar\p$ integral, we see \sac\ is
equivalent to
\eqn\sae{\eqalign{
&\left<\tilde\o^{0,2}\tilde\o^{2,0}\right>_H\cr
&\phantom{aaaa}= \Fr{1}{\hbox{vol}(\CG)}\int
\!\!\CD \! A^\pr\,\CD \! A^\ppr\,
\CD \p\,\CD \bar\p\,\CD \w \cdots
\cr
&\phantom{aaaa}\times\exp\biggl(\cdots +
\Fr{1}{4\pi^2}\int_X \tr\bigl(i\w F^{1,1} +
\p\wedge\bar\p \bigr)\wedge\o
+\Fr{\e}{8\pi^2}\int_X \Fr{\o^2}{2!}\tr\w^2
\biggr)\times
m!(\bar\G\cdot\G)^m
}}
where  $\G\cdot\bar\G=\int_X\o^{0,2}\wedge\o^{2,0}$ denotes the intersection
number\foot{Actually, the above
equation \sae\ should contain a group theoretical factor due to the trace.
Since we are dealing with $SU(2)$ case, the omitted factor is
$\hbox{dim}(SU(2))^m = 3^m$. It seems to us that mathematicians usually
omit this term.}.
Thus we have the following factorization;
\eqn\saf{\eqalign{
\left<(\tilde\o^{0,2}\tilde\o^{2,0})^m\right>_H
&=\sum_{r,s}^{r + 2s=d-2m}\Fr{\e^s}{r!s!}
\left<(\tilde\o^{0,2}\tilde\o^{2,0})^m\tilde\o^r\Theta^s\right>
+\cdots
\cr
&=m!\sum_{r,s}^{r + 2s=d-2m}\Fr{\e^s}{r!s!}
\left<\tilde\o^r\Theta^s\right>(\bar\G\cdot\G)^m
+\cdots,
\cr
}}
that is,
\eqn\sah{
\sum_{r,s}^{r + 2s=d-2m}
\left<(\tilde\o^{0,2}\tilde\o^{2,0})^m\tilde\o^r\Theta^s\right>
=m!\sum_{r,s}^{r + 2s=d-2m}
\left<\tilde\o^r\Theta^s\right>(\bar\G\cdot\G)^m.
}

If $d=2m$, we have
\eqn\sai{
\left<(\tilde\o^{0,2}\tilde\o^{2,0})^m\right>
=m!(\bar\G\cdot\G)^m <1>.
}
Then
\eqn\saj{
\left<(\tilde\o^{0,2}+ \tilde\o^{2,0})^d\right>
=\Fr{2m!}{m!m!}\left<(\tilde\o^{0,2}\tilde\o^{2,0})^m\right>
=\Fr{2m!}{m!}(\bar\G\cdot\G)^m<1>.
}
Equivalently
\eqn\sak{
q_d(\o^{2,0}+\o^{0,2}) = q_0
\Fr{2m!}{m!}\left(\int_X\o^{2,0}\wedge\o^{0,2}\right),
}
where $q_0 =<1>$.
Thus we have
\eqn\sak{
q(\o^{2,0}+\o^{0,2}) = q_0 e^{\int_X\o^{2,0}\wedge\o^{0,2}}.
}

\subsec{General remarks on the algebraic part}

Let $M$ be a {\it simple} simply connected  $4$-manifold with
$b_2^+(M) \geq 3 $.
Let $q_{d}(M)$ denote  the $SU(2)$ polynomials on
$H_0(M,\BZ)\oplus H_2(M,\BZ)$, where $d = 4k -\fr{3}{2}(1 + b^+_2)$.
Kronheimer and Mrowka \KM\ have announced that the Donaldson series
$q(M)=\sum_d q_d(X)/d!$ is given by
\eqn\xxa{
q(M) = e^{Q/2}\sum_{i=1}^{n} a_i e^{K_i},
}
where $Q$ is the intersection form, regarded as a quadratic function
($Q\in \hbox{Sim}^2(H^2(M,\BZ))$), of $M$,
$K_i \in H_2(M)$ denote
the simple classes  and $a_i$ are non-zero rational
numbers.
Since $Q$ is a homeomorphism invariant, any relevant information for
smooth structures is contained in $K_i$ and $a_i$.

Recently, Brusse proved that the basic classes $K_i$ are of
the type $(1,1)$, i.e. $K_i\in H^{1,1}(X,\BZ)$, for a {\it simple} simply
connected algebraic surfaces $X$ with $p_g(X) \geq 1$ \BrB,
using the pureness of the Donaldson invariants
for simply connected algebraic surfaces \BrA.
Then, one of his Corollary
that for all $\o^{0,2} \in H^{0,2}(X,\BZ)$
\eqn\xxb{
q(\o^{0,2} +\o^{2,0}) = q_0 e^{\int \o^{0,2}\wedge\o^{2,0}},
}
where $q_0$ is  Donaldson's polynomial of degree zero,
can be immediately followed from \xxa.
This result says that the algebraic part of Donaldson's polynomials,
i.e.~the polynomials defined by Jun-Li \JL,
contains as much information as the full polynomials for a {\it simple}
simply connected algebraic surface.

More recently, Witten has shown
that all compact K\"{a}hler surfaces
with $p_g \geq 1$ are of simple type \WittenB.
His completely explicit formula for the full polynomials also imply that
all the simple classes (or we should say the Kronheimer-Morwka-Witten
classes) are of the type $(1,1)$, in fact, they are linear
combinations
of components of the canonical divisor \WittenP.

Our heuristic calculation   showes  that
for every simply connected compact K\"{a}hler surface $X$
with $p_g(X)\geq 1$ and for all $\o^{0,2}\in H^{0,2}(X,\BZ)$
\eqn\xxc{
q(\o^{0,2} +\o^{2,0}) = q_0 e^{\int \o^{0,2}\wedge\o^{2,0}}.
}

All of the relevant information (beyond the classical invariants) of
Donaldson's polynomial invariants are contained in the algebraic
part. We should be able to evaluate something like
\eqn\faa{\eqalign{
&\Fr{1}{\hbox{vol}(\CG)}\int
\!\!\CD \! A^\pr\,\CD \! A^\ppr\,
\CD \p\,\CD \bar\p\,\CD \w\,\CD H^{2,0}\,\CD H^{0,2}\,
\CD\c^{2,0}\, \CD \bar\c^{0,2}
\cr
&\times\exp\biggl(
  \Fr{1}{h^2}\!\int_X\!\!\! \tr\biggl[
  iH^{2,0}\wedge *F^{0,2}
  +iH^{0,2}\wedge* F^{2,0}
  -i\c^{2,0}\wedge *\Dpp\bar\p
  -i\bar\c^{0,2}\wedge *\Dp\p
  \biggl]
  \cr
&
+\Fr{1}{4\pi^2}\int_X \tr\bigl(i\w F^{1,1}
+\p\wedge\bar\p \bigr)\wedge\o
+\Fr{\e}{8\pi^2}\int_X \Fr{\o^2}{2!}\tr\w^2
+\Fr{1}{8\pi^2}\int_X \tr(\p\wedge\p)\wedge\o^{0,2}
\biggr)\cr
&\times
\prod_{i=1}^{2m}\Fr{1}{4\pi^2}\int_{\S_i}\tr(i\w F^{1,1} +\p\wedge\bar\p)
,
}}
where $\S_i$ are algebraic cycles Poincar\'{e} dual
to elments of $H^{1,1}(X,\BZ)$. In evaluating this expectation
value, the term $\tilde\o^{2,0}$ itself do not contribute to the
path integral.
However, the modification of the transformation law of $\bar\c^{0,2}$ given
by
Eq.(3.11) dramatically changes the fixed points of the theory.

The path integral of a cohomological field theory with global fermionic
symmetry $Q$ is localized to an $Q$-invariant neighborhood of the fixed
point locus of $Q$. One must perform the path integral along the fixed
point locus exactly, while the transverse path integral can be done in
one-loop approximation \WittenE.

A fixed point of the HYM with action $S_H$ is
\eqn\vbg{
d_A\w =0,
}
where
$A\in\CA^{1,1}$. Thus $\w$ at the fixed point locus is a covariant
constant. There can be two branches;
a) If Eq.\vbg\ has no non-trivial solutions, that is $\w =0$,
the connection $A$
is irreducible.
b) If $\w\neq 0$ solves Eq.\vbg, a holomorphic connection $A$ should be
reducible and the bundle $\CE_A$ splits as a direct sum
of holomorphic line bundles $\CE_A = L\oplus L^{-1}$.
It is worthwhile to note that every higher critical points of HYM theory
are reducible holomorphic connections and there are no reducible
instantons for generic choices of metric in cases of $p_g(X)\geq 1$.
The HYM theory with action $S^\pr_H$ has additional source of fixed
point
\eqn\fab{
\bbs\bar\c^{0,2}= \fr{h^2}{4\pi^2}\w(x)\o^{0,2}(x) =0.
}
Let $C \subset X$ be the locus of $\o^{0,2}(x) = 0$ and $C^c \subset X$
be the locus
of $\o^{0,2}(x)\neq 0$. The equation \fab\ forces that
$\w(x)$ should vanish if $x \in C^c$. On the other hand
$\w(x)$ can be either zero or non-zero covariant constant if
$x \in C$. That is, we have actually three different branches;
i) branch I : If $x\in C^c$, $\w(x) = 0$,
ii) branch IIa : $x\in C$ and $\w(x) = 0$,
iii) branch IIb :$x\in C$ and $\w(x) \neq 0$.
Thus, the path integral \faa\ can be formally written
as product of the contributions $P(I)$ and $P(II)$ of the Branch I
and the Branch II, respectively,
\eqn\yav{
P(I)P(II) = P(I)P(IIa) + P(I)P(IIb).
}

We can evaluate the first part $P(I)P(IIa)$ of the path integral
using the similar method adapted in the previous subsection.
Note that
the higher critical points do not contribute to this path integral,
since $\w = 0$ at the fixed points in Branches I and IIa.
For simplicity, we consider $d = 2m$.
We can simply set $\w=0$ in \faa, which leads to
\eqn\fad{\eqalign{
\Fr{1}{\hbox{vol}(\CG)}\int_{T\CA^{1,1}}
\!\!\CD \p\,\CD \bar\p\,
&\exp\biggl(
\Fr{1}{4\pi^2}\int_X \tr\bigl(\p\wedge\bar\p \bigr)\wedge\o
\biggr) \cr
&\times
\Fr{1}{4\pi^2}\int_{\S_1}\tr(\p\wedge\bar\p) \cdots
\cdot\Fr{1}{4\pi^2}\int_{\S_{2m}}\tr(\p\wedge\bar\p).\cr
}
}
Now the Gaussian integral over $\p$ and $\bar\p$ using \sad\
immediately gives
\eqn\fae{
\Fr{1}{\hbox{vol}(\CG)}\int_{T\CA^{1,1}}
\!\!\CD \p\,\CD \bar\p\,
\exp\biggl(
\Fr{1}{4\pi^2}\int_X \tr\bigl(\p\wedge\bar\p \bigr)\wedge\o
\biggr)Q^{(m)}(\S_1,...,\S_{2m}),
}
where $Q^{(m)}$ is a multi-linear form \DK\ on $H_2(X)$ defined by
\eqn\faf{
Q^{(m)}(\S_1,...,\S_{2m})
=\Fr{1}{2^m m!}\sum_{\s\in S_{2m}}
Q(\S_{\s(1)},\S_{\s(2)})\times\ldots\times
Q(\S_{\s(2m-1)},\S_{\s(2m)}),
}
and $Q$ is the intersection form of $X$.
In particular, if we consider a simply connected hyper-K\"{a}hler
surface such that $\o^{0,2}\in
H^{0,2}(X,\BZ)$ is nowhere vanishing, then, only the branch I
contribute $(C^c \equiv X)$ and we can set $d=2m$.
We have
\eqn\fffi{
q_{2m}(\S_1,...,\S_{2m}) = Q^{(m)}(\S_1,...,\S_{2m}),
}
which coincides to the known mathematical answer \DK\Grady.

\ack{One of us (JSP) would like to thank E.~Witten
for sending him the preliminary version of \WittenB,
which motivated this paper, and for useful communications.
We would like to thank M.~Blau
and G.~Thompson for useful discussions.
}
\parskip =0pt
\listrefs
\bye